\documentclass[grl,draft]{agu2001}

\usepackage{graphicx}
\authorrunninghead{BRAUN, DITLEVSEN AND CHIALVO}

\titlerunninghead{CLIMATE SHIFTS \& LINEAR PHASE RELATIONS}

\authoraddr{H. Braun, Heidelberg Academy of Sciences and Humanities, c/o
  Institute of Environmental Physics, University of Heidelberg, Im Neuenheimer
  Feld 229, 69120 Heidelberg, Germany. (Holger.Braun@iup.uni-heidelberg.de)}

\begin{document}

%
%
%
%
%

%
%

\title{Solar forced Dansgaard-Oeschger events and their phase relation with
  solar proxies}
%
%


\author{H. Braun}
\affil{Heidelberg Academy of Sciences and Humanities,
University of Heidelberg, Heidelberg, Germany}

\author{P. Ditlevsen}
\affil{Niels Bohr Institute, University of Copenhagen,
Copenhagen, Denmark}

\author{D.R. Chialvo}
\affil{Department of Physiology, Feinberg Medical School, Northwestern
  University, Chicago, USA.}

%
%


\begin{abstract}
North Atlantic climate during glacial times was characterized by
large-amplitude switchings, the Dansgaard-Oeschger (DO) events, with an
apparent tendency to recur preferably in multiples of about 1470 years. Recent
work interpreted these intervals as resulting from a subharmonic response of a
highly nonlinear system to quasi-periodic solar forcing plus
noise. This hypothesis was challenged as inconsistent with the observed
variability in the phase relation between proxies of solar activity and
Greenland climate. Here we reject the claim of inconsistency by showing that
this phase variability is a robust, generic feature of the nonlinear
dynamics of DO events, as described by a model. This variability is
expected from the fact that the events are threshold crossing
events, resulting from a cooperative process between the periodic forcing and
the noise. This process produces a fluctuating phase relation with the
periodic forcing, consistent with proxies of solar activity and Greenland climate.   
\end{abstract}

%
%

%

\begin{article}

%
%

\section{Introduction}
Climate archives from the North Atlantic region show repeated
shifts in glacial climate, the Dansgaard-Oeschger (DO) events
\citep{Grootes1993, Andersen2006}. During Marine Isotope Stages 2 and 3 the
intervals between the events exhibit a tendency to coincide approximately with
multiples of 1470 years \citep{Rahmstorf2003}, as depicted in figure 1. The
statistical significance of this pattern and the responsible mechanism,
however, is still a matter of debate \citep{Ditlevsen2005}.
Several hypotheses were proposed to explain the timing of DO
events. One of these relates the events to two century-scale
solar cycles with periods close to 1470/7 (=210) and 1470/17
($\approx$87) years \citep{Braun2005}, the so-called De Vries/Suess
\citep{Wagner2001} and Gleissberg \citep{Peristykh2003}
cycles. Support for a leading solar role comes from deep-sea sediments,
which indicate that during the Holocene century-scale solar variability was
a main driver of multi-centennial scale climate changes in the North
Atlantic region \citep{Bond2001}.  

\begin{figure}
\noindent
\includegraphics[width=20pc,height=15pc]{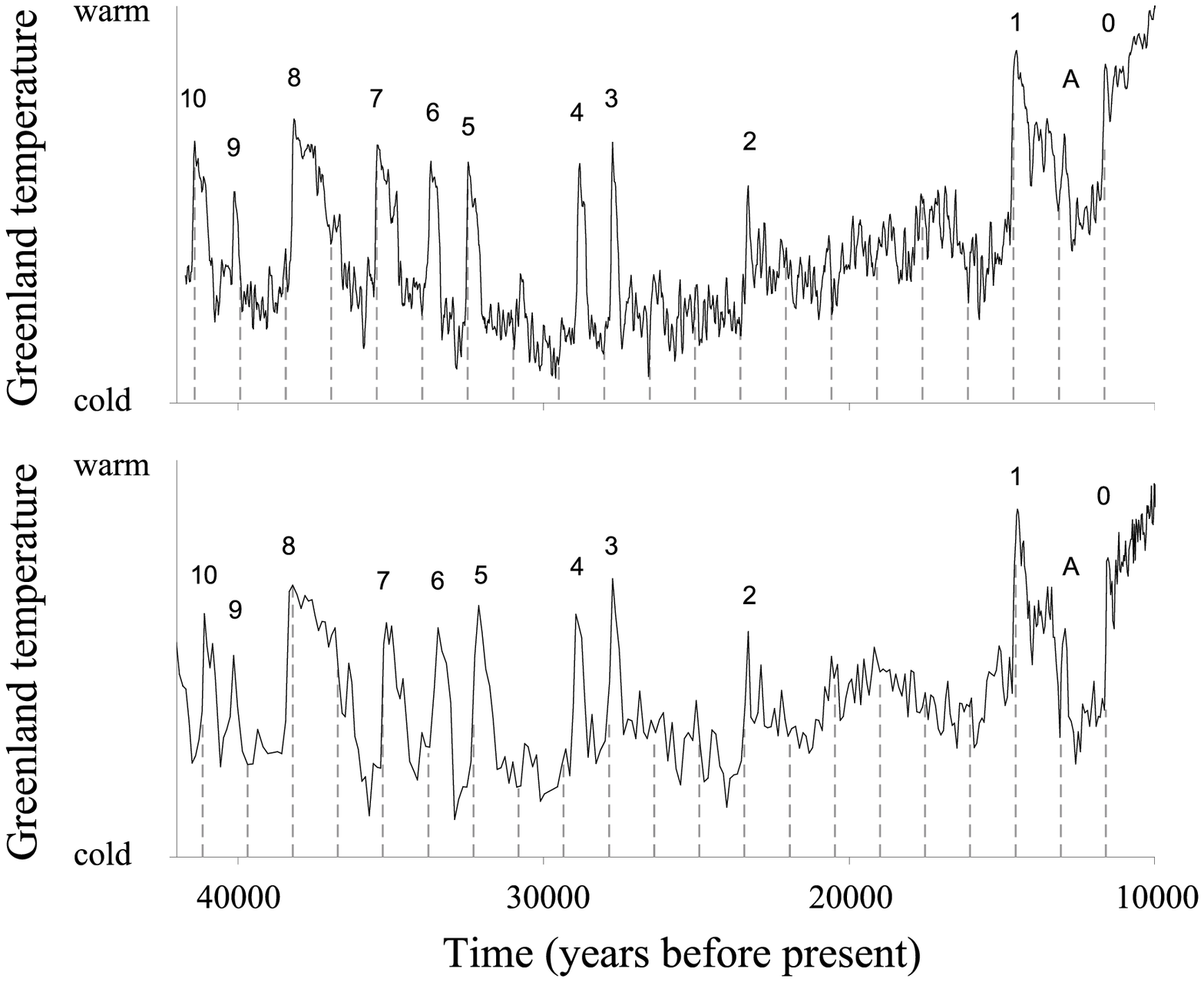}
 \caption{DO events as seen in two ice cores from Greenland. Top:
   NGRIP. Bottom: GISP2. Labels mark the events 1-10, the Aller{\o}d (A) and
   the end of the Younger Dryas (0). Dashed lines are spaced by
   1470 years.}
\end{figure}

Recently the phase relation between solar variability (deduced
from $^{10}$Be) and 14 DO events was analyzed \citep{Muscheler2006}. A
relation far from fixed was found and was interpreted as being in
contradiction to Braun et al.'s hypothesis \citep{Braun2005}. While in linear systems a
constant phase relation between the forcing and the response
is expected, such a relation does not necessarily exist in
non-linear systems. But climate records and ocean-atmosphere models
\citep{Ganopolski2001}, which are not yet suitable for statistical analyses on
DO events because of their large computational cost, suggest that the events
represent switches between two climate states, consequently implying an
intrinsically non-linear dynamical scenario. Thus, to interpret the reported
lack of a fixed phase relation between the DO events and solar proxies, it is
crucial to analyze their phase relation in simple models.


%
%

\section{A Simple Model of DO Events}


%
%

Here we investigate this phase relationship in a very simple model of DO
events. A comprehensive description of this model has already been published
before, including a detailed discussion of its geophysical motivation and its
applicability, as well as a 
comparison with a much more detailed ocean-atmosphere model \citep{Braun2007}.
In the simple model, which was derived from the dynamics of the
events in that ocean-atmosphere model, DO events represent repeated
switches between two possible states of operation of a bistable,
excitable system with a threshold (Figure 2). These states
correspond to cold and warm periods of the North Atlantic region
during DO cycles. The switches are assumed to occur each time the
forcing function (f), which mimics the solar role in driving DO events,
crosses the threshold (T). Transitions between the two states are accompanied
by an overshooting of the threshold, after which the system relaxes back to
its respective equilibrium following a millennial-scale relaxation. 

The rules for the transitions between both states are illustrated
in figure 2. It is assumed that the threshold function $T$ is
positive in the interstadial (``warm'') state and negative in the
stadial (``cold'') state. A switch from the stadial state to the
interstadial one is triggered when the forcing $f$ is smaller than
the threshold function, i.e. when $f(t)<T(t)$. The opposite switch
occurs when $f(t)>T(t)$. During the switches a discontinuity in
the threshold function is assumed, i.e. T overshoots and takes a
non-equilibrium value ($A_{0}$ during the shift into the stadial
state, $A_{1}$ during the opposite shift). Afterwards, T
approaches its new equilibrium value ($B_0$ in the stadial state,
$B_1$ in the interstadial state) following a millennial scale
relaxation process:
\begin{equation}
{\frac{dT}{dt} = -\frac{T-B_{s}}{\tau_{s}}.}
\end{equation}
Here, $\tau_0$ and $\tau_1$ represent the relaxation time in the
stadial state (s=0) and in the interstadial state (s=1), respectively.

Both the overshooting relaxation assumption and the transition
rules in our simple model are a first order approximation of the
dynamics of DO events in a coupled ocean-atmosphere model
\citep{Ganopolski2001}. In that model the events also represent
threshold-like switches  in a system with two possible states of
operation (corresponding to two fundamentally different modes of
deep water formation in the North Atlantic) and with an
overshooting in the stability of the system during these shifts
\citep{Ganopolski2001, Braun2007}. Analogous to the simple model,
switches from the stadial mode into the interstadial one are
triggered by sufficiently large negative forcing anomalies (i.e.
by a reduction in the surface freshwater flux to the North
Atlantic that exceeds a certain threshold value), whereas the
opposite shifts are triggered by sufficiently large positive
forcing anomalies (i.e. by an increase in the freshwater flux that
exceeds a certain threshold value). It has further been
demonstrated that the simple model is able to reproduce the timing
of DO events as simulated with the ocean-atmosphere model, as well
as the occurrence of non-linear resonance phenomena such as
stochastic resonance and ghost resonance, which were shown to be
properties exhibited by that model \citep{Ganopolski2002,
Braun2005, Braun2007}.

An obvious advantage of the conceptual model compared with the
ocean-atmosphere model is its low computational cost, which allows
for extensive statistical analyses on the timing of DO
events. All model-parameter values chosen here are the same as in two earlier
publications (A$_0=-27$ mSv, A$_1=27$ mSv, B$_0=-9.7$
mSv, B$_1=11.2$ mSv,
 $\tau_0=1200$ years, $\tau_1=800$ years; 1 mSv = 1
milli-Sverdrup
 = $10^{3}$ m$^{3}$/s) \citep{Braun2005, Braun2007}.

\begin{figure}
\noindent
\includegraphics[width=18pc,height=14pc]{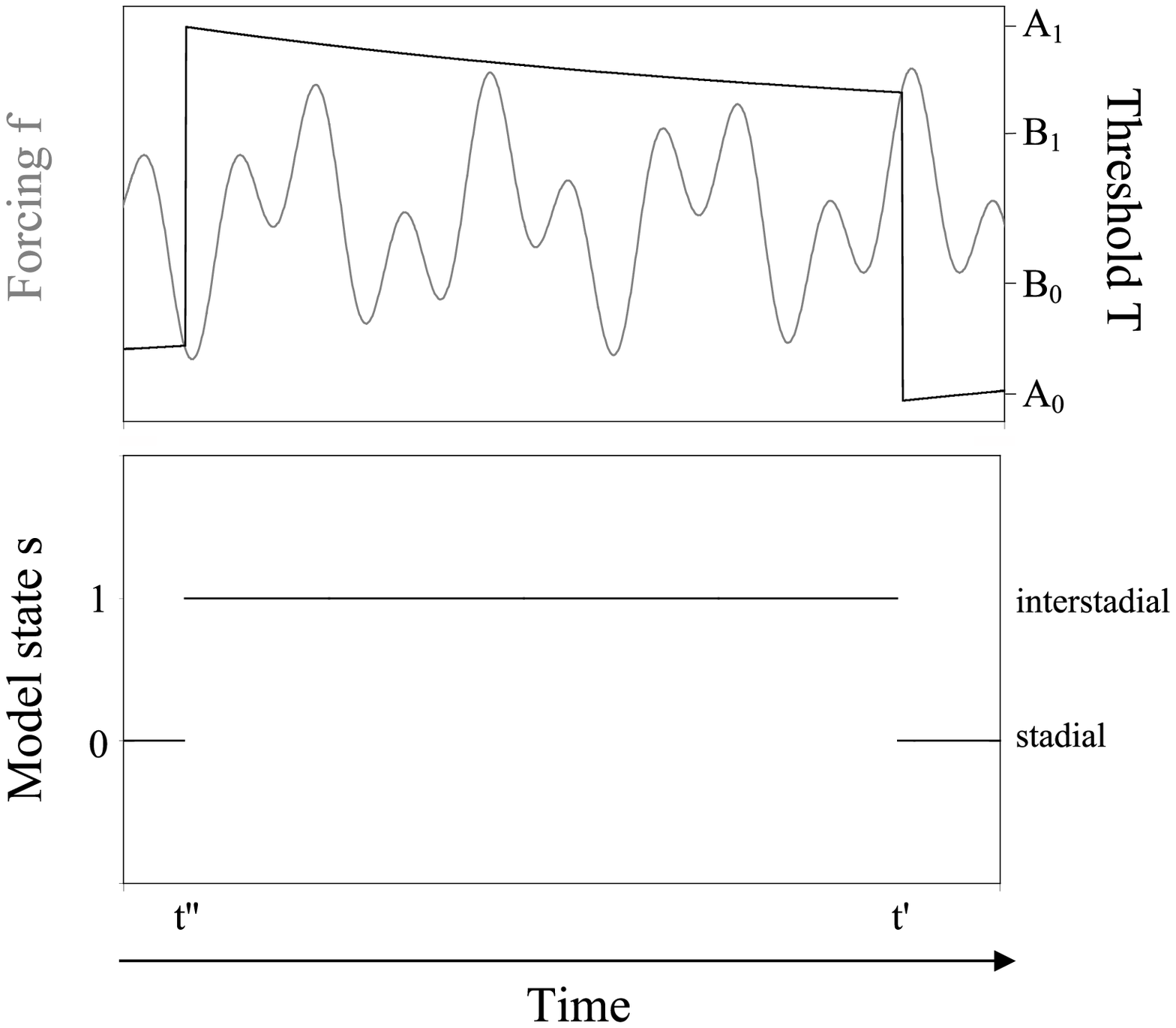}
 \caption{Dynamics of DO events in the model. Top: Forcing f (grey)
   and threshold function T (black). Bottom: Model state s (s=0: ``cold''
   state, $s=1$: ``warm'' state). A switch from the cold to the warm
   state is triggered when $f<T$, which happens in this example at time $t''$.
   During the transition, interpreted as the start of a DO event, $T$
   overshoots and relaxes back towards its new equilibrium $B_{1}$ ($B_{1}>0$)
following a millennial time scale. The events are terminated by a switch back
to the cold state, which is triggered when
   $f>T$ (at time $t'$ in the figure). Again, T overshoots 
   and approaches its new equilibrium $B_{0}$ ($B_{0}<0$).}
\end{figure}

\section{Testing Fixed Phase Relationships}

To test the assumption of a fixed phase relationship between
solar-forced DO events and solar variability, we here drive our model
by a simple input consisting of noise $\sigma \cdot n(t)$ and of two
sinusoidal cycles with equal amplitudes A:
\begin{equation}
{f(t) = -A \cdot ( \cos[\frac{2 \pi t}{T_1}] + \cos[\frac{2 \pi t}{T_2}]) +
  \sigma \cdot n(t).}
\end{equation}
 $\sigma$ is the standard deviation
of the noise and $n(t)$ the standard unit variance white noise, with a cutoff
frequency of 1/50 years (figure 3).
Following \citet{Braun2007} the cutoff is
used to account for the fact that the model shows an unrealistically large
sensitivity to decadal-scale or faster forcing.
In analogy to \citet{Braun2005} the periods of the two cycles are chosen to be $T_{1}$ =
1470/7 (=210) years and $T_{2}$ = 1470/17 ($\approx$86.5) years, i.e. close to
the leading spectral components of the solar De Vries and Gleissberg
cycles.

In the simulations shown in Figure 4 we use three
different signal-to-noise ratios (SNR): A = 8 mSv and $\sigma$ =
5.5 mSv ($SNR \approx 2.1$), A = 5 mSv and $\sigma$ = 8 mSv ($SNR
\approx 0.4$), A = 3 mSv and $\sigma$ = 9 mSv ($SNR \approx 0.1$).
For all of these, the waiting time distribution of the simulated
events is centered around a value of 1470 years, with several peaks of only
decadal-scale width (figure 4). The relative 
position of these peaks is well understood in the context of the
ghost resonance theory \citep{Chialvo2002, Chialvo2003, Calvo2006,
  Braun2007}. The peaks result from constructive interference between the two
sinusoidal forcing cycles which produces particularly large magnitude
variations in the bi-sinusoidal forcing and -- when noise is added --
leads to favored transitions at the corresponding waiting times. Depending on
the relative amplitude values of the noise
and the periodic forcing this synchronization is more or less efficient,
as is seen for the different signal-to-noise ratios in Figure 4. Even for the
lowest ratio, however, the synchronization is
still notable. The waiting time distributions shown in figure 4 are thus almost
symmetrically centered around a preferred value of 1470 years because the
sinusoidal cycles enter in phase every 1470 years, creating forcing peaks of
particularly large magnitude. This 1470-year repeated
coincidence of the bi-sinusoidal forcing, however, does not show
up as a corresponding forcing frequency, since no sinusoidal cycle
with that period is present. Thus, when linear spectral analysis is performed on
the forcing, only the two century-scale sinusoidal cycles are detected as
outstanding components.

\begin{figure}
\noindent
\includegraphics[width=20pc,height=14pc]{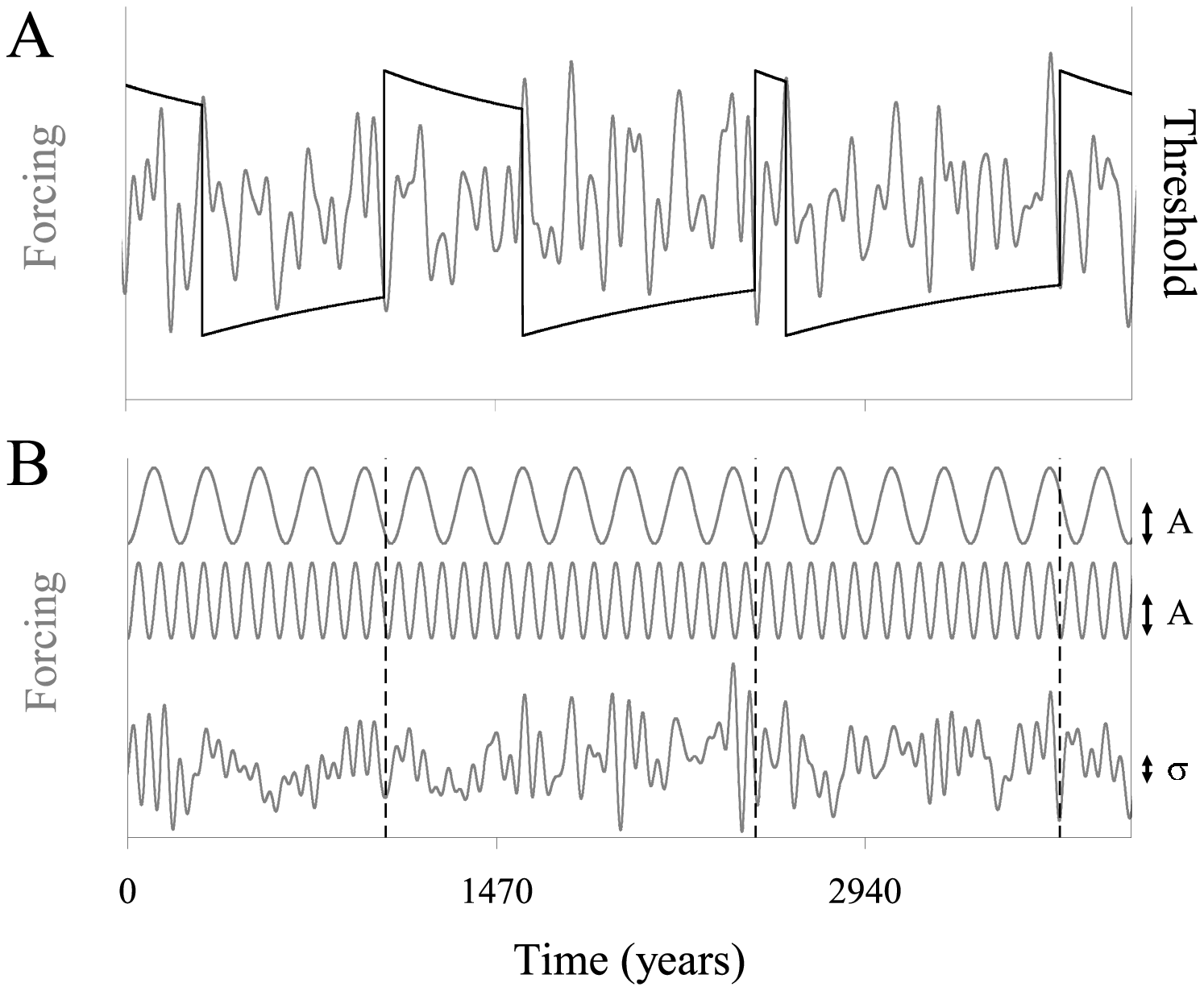}
 \caption{Forcing and response. The forcing consists of two
  added sinusoidal cycles with equal amplitudes ($A=8$ mSv) and white noise
   ($\sigma=5.5$ mSv). (A) Total forcing f (grey) and threshold function T
   (black). (B) Forcing components
   (grey); from top to bottom: 210-year cycle, 86.5-year cycle, noise. Dashed
   lines indicate the onset of the simulated DO events. Despite the tendency
   of the three events to recur approximately every 1470 
   years, only the first two events coincide with minima of the 210-year
   cycle. The third event, in contrast, occurs
   closely after a maximum of that cycle.}
\end{figure}

\begin{figure*}
\noindent
\includegraphics[width=39pc,height=18pc]{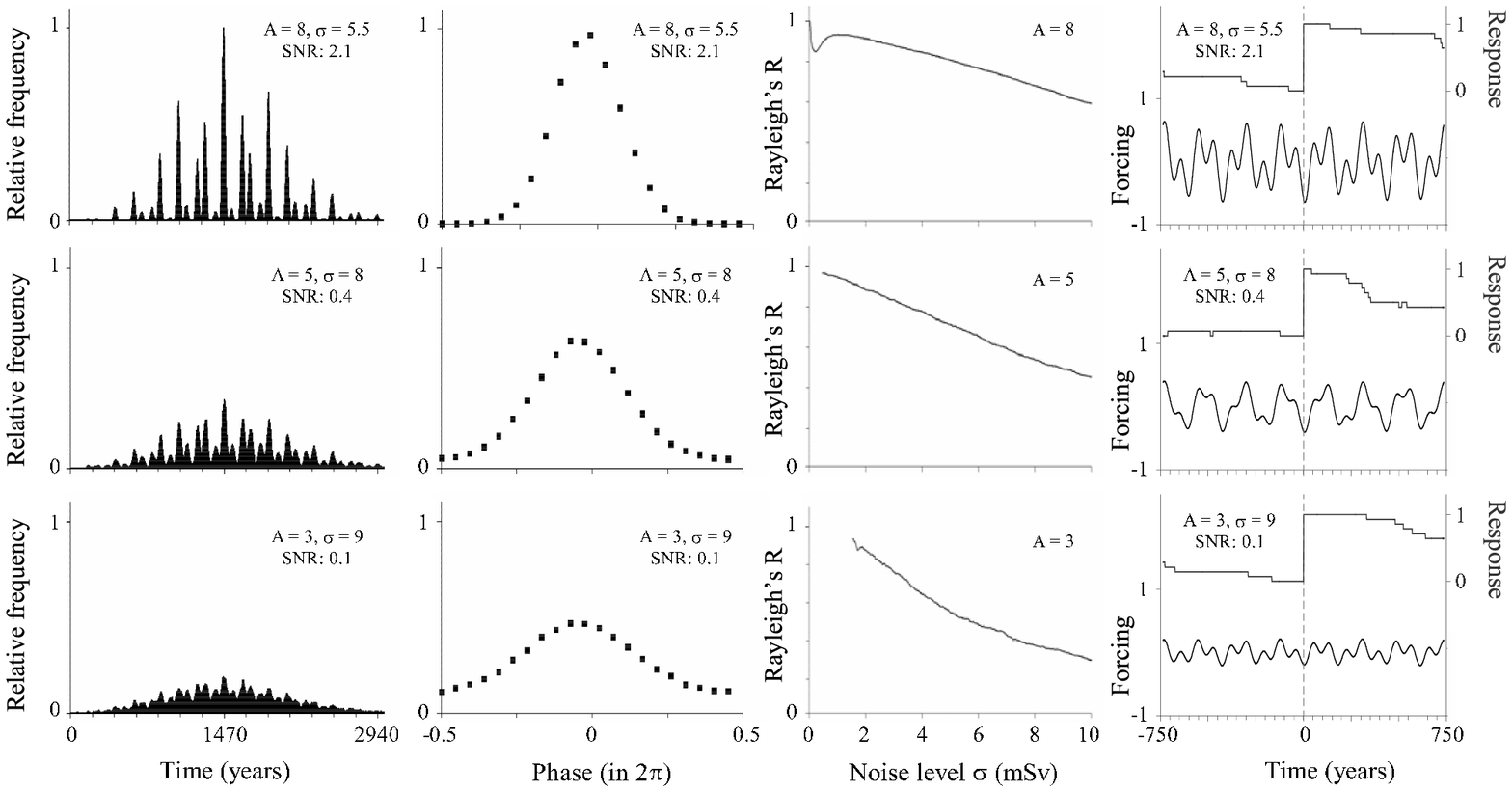}
 \caption{Waiting time distribution of the simulated events and their phase
   relation with the forcing. The amplitude of the two
   sinusoidal forcing cycles and the standard deviation of the noise
   are: A = 8 mSv, $\sigma$ = 5.5 mSv (top); A = 5 mSv, $\sigma$ = 8 mSv
   (middle); A = 3 mSv, $\sigma$ = 9 mSv (bottom). First column: normalized
   distribution of the spacing between successive DO events in the
   simulation. Second column: probability distribution of the
   phase relation between the onset of the events and the
   210-year cycle (zero corresponds to a start of the events at the
   minimum of that cycle, $\pm \pi$ to a start at the maximum). Third
   column: Rayleigh's R value \citep{Ditlevsen2007}, a measure for the
   phase correlation between the forcing cycles and the simulated events,
   as a function of the noise level $\sigma$ ($R = \frac{1}{N} | \sum_{n=1}^{N} \cos \frac{2
     \pi t_n}{T_1} + i \cdot \sin \frac{2 \pi t_n}{T_1} | $, where $t_n$
   denotes the timing of the events, $T_1$ = 210 years and $N \to
   \infty$). R is 1 if and only if a fixed phase relation exists between
   the 210-year forcing cycle and the events. Fourth column: Superposition of
   the model response (i.e. of the state variable s) and of the bi-sinusoidal
   forcing over a series of 14 simulated events, aligned by the onset of the
   events following \citet{Muscheler2006}. The
   unaveraged bi-sinusoidal forcing is normalized with maximum and minimum values of $\pm$1.} 
\end{figure*}

Despite the robustness of the synchronization effect, none of the two
sinusoidal cycles in our forcing shows a fixed phase relationship with {\it{all}} of
the simulated DO events, due to the presence of noise and the existence of
a threshold. In our model, a fixed phase relation can only be present in the low
noise limit (i.e. either for $\sigma \to 0$ [with a supra-threshold
bi-sinusoidal forcing] or for the lowest noise level that still enables
repeated threshold crossings [with a sub-threshold bi-sinusoidal forcing],
thus corresponding to DO events with extremely long waiting times), compare
third column in
figure 4. Even for the largest of the three signal-to-noise ratios in our simulations, the events thus only show a
tendency to cluster around a preferred phase of the two sinusoidal
cycles. Outliers, however, can still occur in almost opposite phase, at least once over a sufficiently
large number of events in the simulation. For the highest signal-to-noise
ratio, for example, there is still a probability of about 35 percent to find
at least one out of 14 events in opposite phase (i.e. outside of the
interval $[-\pi/2,+\pi/2]$). And for the other two signal-to-noise ratios, the
corresponding probability is even much higher (i.e. 92 percent and 99.5
percent, respectively). Since a fixed phase relationship between the simulated
events and the forcing cycles does not even exist in our very simple model
system, it appears unrealistic to us to assume the existence of such a
relationship in the climate system. Thus, the reported lack of a fixed phase
relationship between DO events and solar variability (deduced from $^{10}$Be)
would also be expected with the proposed ghost resonance solar forcing. We note that
superimposed epoch analyses of 14 simulated events can produce forcing-response
relations similar to the one reported by \citet{Muscheler2006}: The onset of
the superimposed events (at $t=0$ in the fourth column in figure 4) typically
coincides with a minimum in the averaged bi-sinusoidal forcing which, however,
is not more pronounced than other minima and is highly damped as compared with
the unaveraged forcing. A considerable statistical spread exists in the
magnitude of this damping because the small number (14) of events is not yet
sufficient to infer reliable information concerning the average phase
fluctuation between the input and the output.

This lack of phase correlation between forcing and response is explained by
the threshold character of DO events: The simulated events are triggered when
the total forcing
(the sum of the two sinusoidal cycles and the noise) crosses the
threshold function. Some of the threshold-crossings are in the first place
caused by constructive interference of both cycles. These events
coincide with near-minima of the two forcing cycles. Other
threshold-crossings are, however, in the first place caused by constructive
interference of just one cycle and noise. These events thus
coincide with a near-minimum of only that cycle (compare figure 3), whereas a fixed
phase-relation with the second cycle does not necessarily exist. And at least
for low signal-to-noise ratios, some of the threshold crossings are in the first
place caused by the noise alone. These events thus do not show a fixed phase
relation with any of the two forcing cycles.

The inherently nonlinear noisy synchronization mechanism exhibited
by our model is not unique to DO events. In fact, it
has originally been proposed to explain the perception of the
pitch of complex sounds \citep{Chialvo2002, Chialvo2003} and, as a
general concept, has already been used to describe theoretically
and experimentally similar dynamics in other
excitable \citep{Calvo2006} or multi-stable systems with thresholds,
e.g. in lasers \citep{Chialvo2002, Chialvo2003, Buldu2003}.
Because of the fact that the leading output frequency is
absent in the input, this type of resonance is called ghost
stochastic resonance.

\section{Conclusions}

We here used a simple model of DO events, driven by a bi-sinusoidal forcing
plus noise, to show that a fixed phase relation between the forcing cycles
and {\it{all}} simulated events does not exist, apart from the unrealistic low
noise limit. As argued above, in this model the fluctuations in the
phases between the forcing and the response are related to the
process giving rise to the transition itself. Each event is
generated by a threshold crossing resulting from a cooperative
process between the two periodic driving forces (i.e., the
centennial-scale input cycles) and the stochastic fluctuations. In
this nonlinear scenario, as we showed explicitly in our simulations,
millennial-scale events with fixed input-output phase relations are impossible for any nonzero noise amplitude.

While one could disagree on the interpretation and the
statistical significance of the pattern described in Figure 1, our
results show that the reported lack of a fixed phase
relationship between 14 DO events and solar proxies is consistent
with the suggested solar role in synchronizing DO events. At the
same time our results have further implications for a second so-far
unexplained oscillation during Pleistocene climate, i.e. the
glacial-interglacial cycles, which also show strong indications for the
existence of threshold-like dynamics during glacial terminations
\citep{Paillard1998, Huybers2005}: Since the existence of a causal
relation between threshold-crossing events and their quasi-periodic
forcing does not necessarily imply the existence of a clear phase relation
over {\it{all}} events, the lack of such a relation between glacial
terminations on one hand and the orbital eccentricity and precession cycles on
the other hand is not sufficient to reject a leading role of these cycles
during terminations, in contrast to the interpretation proposed by
\citet{Huybers2005}. More insight in the cause of Pleistocene climate cycles
might thus be gained from more adequate statistical approaches, based e.g. on
Monte-Carlo simulations \citep{Ditlevsen2007} with simple models
\citep{Paillard1998, Braun2007} that mimic the nonlinear dynamics which seems
to be relevant during these oscillations.

\begin{acknowledgments}
H.B. thanks A. Ganopolski for instructive discussions on the dynamics of DO
events and A. Svensson for his hospitality at the Niels Bohr
Institute. H. B. was funded by Deutsche Forschungsgemeinschaft (project MA 821/33).
\end{acknowledgments}

\end{article}

\end{document}